**Population Growth and Economic Development in Bangladesh: Revisited Malthus**

Md Niaz Murshed Chowdhury[1]
Department of Economics
University of Nevada Reno

Md. Mobarak Hossain[2]
Department of Economics
University of Nevada Reno

**Abstract**

Bangladesh is the 2nd largest growing country in the world in 2016 with 7.1% GDP growth. This study undertakes an econometric analysis to examine the relationship between population growth and economic development. The result indicates population growth adversely related to per capita GDP growth, which means rapid population growth is a real problem for the development of Bangladesh. Malthus's prediction is that population increases so rapidly and outstrip the food supply due to the operation of the law of diminishing return, which is proven wrong because of technological improvement, human capital development and so on in Bangladesh. Bangladesh has reduced its population growth by about 67% between 1979 and 2017 using different preventive checks suggested by Malthus and Mill. Bangladesh has been suffering from environmental degradation, loss of arable land, loos of agricultural land biodiversity loss and deforestation. As a consequence climate has changed dramatically and species are in danger and extinction.

---

[1] Corresponding Author, University of Nevada Reno, Department of Economics, Reno USA; Email: mdniazc@nevada.unr.edu
[2] University of Nevada Reno, Department of Economics, Reno USA. Email: mhossain@unr.edu

**Introduction**

Bangladesh is one of the most densely populated countries in the world with 3310 people per square mile (1278 per Km$^2$) where the US is only 92.2 per square mile. The population density of Bangladesh is three times higher than neighboring country India and 36 times higher than the United States. According to the report of the United Nations Population Fund (UNFPA), the Bangladesh population has reached 167.9 million in 2018. Thomas Malthus said unchecked population growth can causes danger in the economy but Bangladesh demography tells a different story. Most of the underdeveloped countries have fallen into the demographic trap due to a high birth rate where Bangladesh has achieved extraordinary success in that area. Bangladesh has been undergoing a population transition in the line with the population transition model and Bangladesh is now at the beginning of stage III of its population transition. In recent decades dead rate decrease substantially where birth rate doesn't fall enough to maintain a sustainable growth of population. The population growth rate in Bangladesh is comparatively higher than other developing countries in late 1970. Population growth plays an intricate, complex and interesting role in the process of the development of any country, sometimes it boosts up the economic development and sometimes it retards the economic development.

Robert Malthus is popular for his famous essay on "Principle of Population" and the most well-known theory of the population is the Malthusian theory. He wrote it in 1798 to warn his fellow countrymen of impending disaster because of increasing population in England. According to Malthus " By nature human food increase in a slow arithmetical ration; man himself increase in a quick geometrical ratio unless want and vice stop him. The increase in numbers is necessarily limited by the means of subsistence Population invariably increase when the means of subsistence increase, unless prevented by powerful and obvious checks."

Malthus population theory is based on the two assumptions, diminishing returns to labor, and the rate of population growth to which the means of subsistence is over the minimum level. Malthus said, population increase faster than food production, population increase in a geometric rate unless prevented by powerful check, and food production increase only in an arithmetic rate. It indicates per head food tends to decrease as the population increases, which means the laws of decreasing returns to labor. He talked about two types of a check to keep the means of subsistence on a level, and these are the preventive check and positive check. Malthus was different from the other classical economist for his "Principle of Political Economy' which deals with progressive of wealth. Malthus talked about Economic development and he emphasized on the short-term development instead of long-term development. In that book, Malthus says " There is scarcely any inquiry more curious or from its importance more worthy of our attention, than that which traces the causes which practically check the progress of wealth in different countries." This paper tries to examine some of his thinking in the perspective of Bangladesh. The key objectives of this paper are, explain the relationship between population growth and per capita GDP growth rate, and summarize different problem that Thomas Malthus's addresses his published books in a sense of Economics Development and explain why some of his dire predictions have not occurred in Bangladesh, and finally explain the concept of demographic translations.

**Population Growths and Economic Development**

The process of population growth is exogenous in order to process income generation, accumulation, technical progress and institutional change (Srinivasan, 1987). The relationship between population growth and economic development has been the main concern in the recent decade. Thomas Malthus argued that population growth would depress living standard in the long run. Excessive population growth could reduce per headland, which pushed downward pressure to the fixed amount of land. Population growth. Robert MacNamara, past president of word Bank, described this relationship as "this most delicate and difficult issue in the era, it is controversial, subtle and immeasurably complex". This study tries to show the relationship between population growth and per capita GDP in this section.

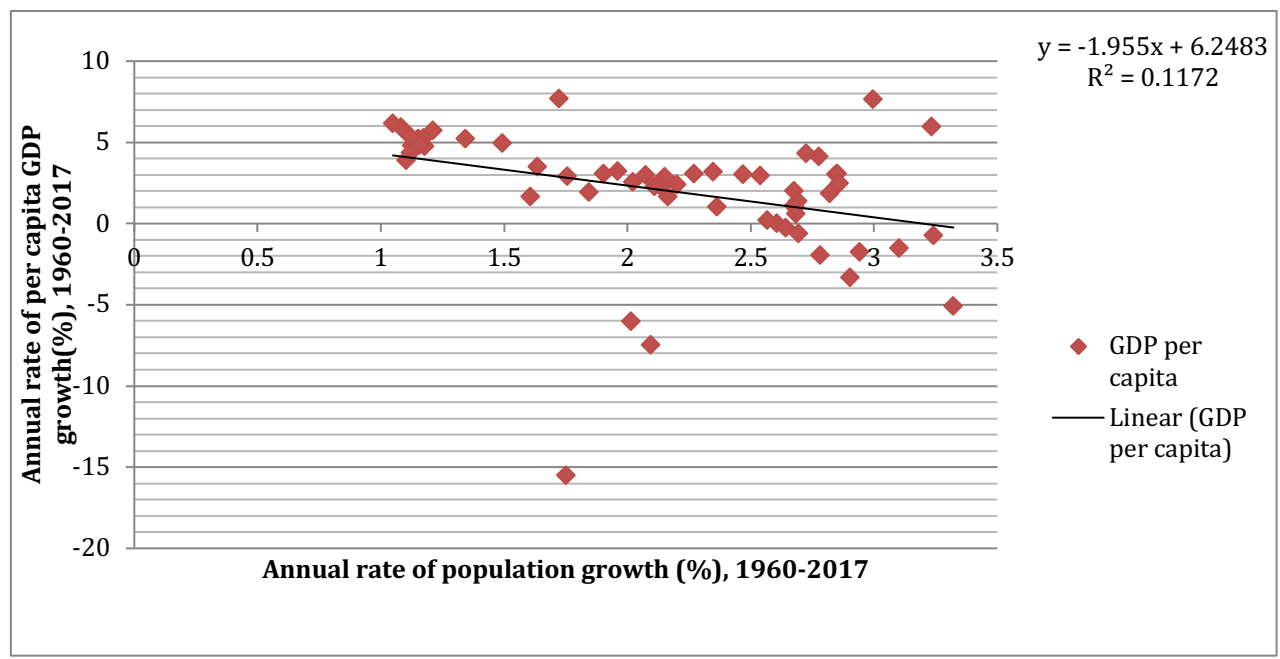

Figure 1: A scatter chart of population growth rates versus GDP per capita growth rates for various developing countries for the period 1960-2017. Source: World Bank Data Bank

Figure 1 demonstrates the simple cross-sectional relationship between population growth and economic growth, which shows a negative correlation when considered over the long run (1960-2017). In the recent decade, Bangladesh achieved great success in reducing its population growth rate, from 3.00% in 1979 to 1% in 2017. It indicates Bangladesh has taken preventive check using community-based approach, recruiting married village women trained in basic medicine and family planning that consists of different birth control methods like contraceptive pills and condoms and referring women for clinical contraception. Bangladesh government has taken steps to prioritized girls education, continuing education delays marriage and educated women have opportunities to work outside. In the process, women education and empowerment helps to reduce fertility rates in Bangladesh subsequently halved from six children in the period 1970 to three children in 2017. Every highly developed country has experienced a reduction in fertility of 50% or more, and a population in which fertility was being reduced would be healthier, better fed and better educated (Zaidan, 1969).

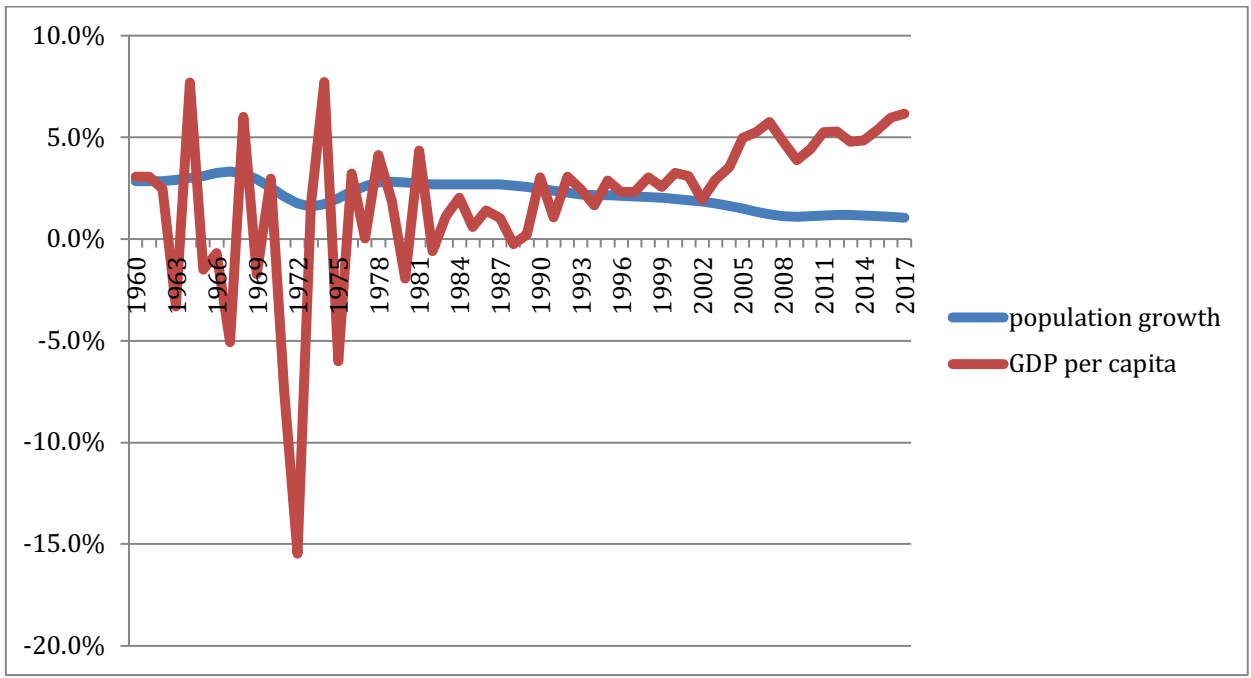

Figure 2: A line chart of population growth rates versus GDP per capita growth rates for Bangladesh for the period 1960 to 2017. Source: World Bank Data Bank

Above Figure 2 provide the graphical representation of per capita Gross domestic product and population growth rate, which shows population growth decrease over time but GDP per capita increases between these periods. We clearly have seen that rapid economic growth mitigates the potential negative impact of rapid population growth. Above diagram shows that Bangladesh Economy was fall about 16% in 1972 due the liberation war against West Pakistan. After that, until 1990 GDP growth was lower than population growth that demonstrates Bangladesh went through severe population problems. After 1990 Bangladesh economy has been flourishing, and Bangladesh has taken control over population growth and reduced population growth substantially from 2.8% in 1990 to 1% in 2017. Even though Bangladesh has taken control over population growth, Bangladesh still suffering from its increased population because they already damaged agricultural land, forestland, and biodiversity.

This study used yearly times series data from the period of 1960 and 2017. For the simplicity of the model[3] the study used only two series of data[4], which are per capita GDP (%) and Population Growth rate (%). A simple linear regression analysis was undertaken to determine the relationship between population growth and economic development.

$$\text{GDP} = 6.25 - 1.95 \, \text{PopulationGrowthRate} \; ; R^2 = 0.1172$$
$$\quad\;\; (1.636)\text{***}[5] \quad (0.716)\text{***}$$

This result indicates there is a clear negative and statistically significant relationship between per capita GDP and population growth rate when we considered over the long run (1960-2017), which indicates 1% increase in population growth leads to decrease per capita GDP by 1.95%. The clear indication from the output is that population growth adversely affects the economic development of Bangladesh, which supports the Malthus prediction. Rapid population growth indeed a real problem for Bangladesh, high population growth has become a great obstacle to achieving overall development goal. Population growth causes a huge problem in Bangladesh, which put pressure on investment growth and diminishes the savings rate (Ali, Alam, Islam, & Hossain, 2015). Millions of Bangladeshi is working abroad and remittances are an important source of foreign exchange earnings, which is the important part of current account balances. In Bangladesh, natural resources are diverted to fulfill consumption needs that put the downward pressure to the national savings rate. Bangladesh had to seek external financial assistance from abroad every year to fulfill the demand of food besides its tremendous success agriculture sector. For these reasons, Bangladesh is one of the lowest domestic savings countries in Asia, 25.3% in Bangladesh compared to 41% in China. Every year Bangladesh is importing huge volume of food from foreign countries, Brazil is the top on that list (imported 57% of foods from Brazil). In order to protect domestic market, high savings rate is necessary for the developing countries because domestic savings in the key sources for domestic investment to induce economic development, reduce dependency from foreign aid and save huge interest payment[6].

---

[3], Some of the previous studies used different models, including real GDP, export, import, mortality rate and so on, but everyone found a negative relationship between population growth and GDP growth rate.
[4] Data was collected from World Bank databank, world development indicator database
[5] Standard errors in parentheses, Significance at the 5% level indicates by ***

**Demographic Transition, population growth, and food production**

"Essay on the Principle of Population" was one of most enduring work during the time of Malthus and growth of population will inevitably collide with diminishing return was the root argument of Malthus. His diminishing return concept doesn't work in recent time in developing countries like Bangladesh because of advanced technology, human capital development, and agricultural revolution. Recent statistics show that the increase in food production surpasses the increase in population growth in Bangladesh. The new invention and advanced methods of production have disproved the forecast of Malthus. Based on a report of the World Bank, Bangladesh accounted for 90% of the reduction in poverty between the periods of 2005 and 2010. This report indicates agriculture food increase substantially in Bangladesh to protect a large number of population in Bangladesh. Agriculture is the key driving force in Bangladesh, more the 70% population and 77% of its works force live in rural areas, 87% of rural household rely on agriculture for at least part of their income. Despite the frequent natural disaster and population growth, Bangladesh has made exemplary progress over the past 40 years in achieving food security due to the public investment in agricultural technology, rural infrastructure, and human capital development. Bangladesh agriculture sector has gained tremendous success in grain production, which triples between 1972 and 2014 (9.8 to 34.4 million tons). Bangladesh is one of the fastest rates of productivity growth in the world since 1995 averaging 2.7% per year (second only to China) and 4.6% agricultural growth between 2010 and 2014. Bangladesh

---

[6] Most of the developing countries like Bangladesh have to pay huge interest for the foreign debt and foreign assistance. If they able to borrow from a local market, they can save huge interest payment and can promote the local investors to invest in the domestic market.

GDP for agriculture increases 30% from 2008 to 2017 (7729 million to 10117 million), which indicates food production doesn't increase arithmetic progression rather geometric progression. Malthus made a mistake when discussing the population question, he took account only agricultural land and food production alone. He should have considered all types of products like natural resources, human capital development, and the export of skilled labor, factories, industries, machinery, mines and other specialized industries. For example, readymade garments sectors in Bangladesh is one of the most influential sectors in Bangladesh, the textile and clothing industries playing a key role to boost the economy of Bangladesh. Export in Manpower and textile and garments are the principal source of foreign exchange earnings for Bangladesh. Bangladesh imported foodstuffs from foreign countries using the foreign exchange reserve to reduce the food problem, and recent statistics show that Bangladesh has reduced the problem of food and increases per capita GDP (PPP) in recent times, which clearly indicates the higher living standard even though the high volume of population pressure.

**Population growth and environmental degradation**

Malthus assumed that excess population causes disaster for the country if the population is not checked. This statement is true for the country like Bangladesh. There is a complex relationship between population and the environment. The excess population causes environmental degradation in many ways. It puts the downward pressure to arable land, agricultural land and forestland that is needed to balance the biodiversity. In Bangladesh, Agricultural land decreases by 13.5% and arable land decrease by 10% between the period of 1989 and 2016, which demonstrate that excess population taking over agricultural land rapidly. Forest areas are decreasing rapidly in Bangladesh that is accounted 4.5% between the period of 1990 and 2016. Recent statistics show that forest area as a percent of land area is 10% in 2017 but according to global forest policy for a country, it should be at least 25% of total land area. Bangladesh is lying far below the standard level and this is one of the root causes of climate change in Bangladesh. In addition, Bangladesh is one of the most disaster-prone countries in the world like floods, cyclones and droughts and everyone these disaster made huge losses in the country. The weather pattern has changed dramatically in Bangladesh, the emission rate of $CO_2$ and led in the air of Bangladesh is very high; consequently, it causes different health problem like cancer, skin diseases and so on. Bangladesh is suffering extreme temperature during summer and excessive rain in the rainy reason due to deforestation. These types of climate change and natural disasters ruin agricultural productivity every year.

**Benefit from population growth**

An increasing population means an increase in the labor force that can participate directly to the development process and economic growth, and a growing population leads to an increase in total output. In addition, a growing population can provide a growing market for most good and services that stimulate entrepreneurs to invest more and more in capital goods and machinery as a consequent business activity will be increased and more income and employment will be generated in the process. In addition, an increasing population can provide cheap labor for the industries and produce export goods at low cost, which is needed for economic development.

Bangladesh is getting several benefits from population growth, for instance, Bangladesh exporting skilled labor forces to different countries, especially oil-based Middle East countries. These labors send remittance to Bangladesh that increases foreign exchange reserve, which is needed for trade balances. Bangladesh has taken several projects to train unskilled labor, and turn them to productive labor by proving appropriate knowledge. Due to the availability of cheap labor foreign investors are encouraged to invest in Bangladesh, in the way Bangladesh gained notable success in textile and readymade garments industry. Bangladesh ranked 4[th] largest clothing industries in the world. It creates huge employment in the root level of Bangladesh, which reduces the unemployment rate and boosts up the living standard and reduces the poverty rate. In addition, it induces capital inflow in Bangladesh through foreign direct investment (FDI) that diminish the capital shortage in Bangladesh. Industries of Bangladesh are enjoying economies of scale in the production process because of the availability of cheap labor. Trade openness has a long-term positive effect on less developed courtiers (Kentor, 2000).

**Malthus, Effective demand and Savings**

Malthus pointed out that the process of development is not automatic; the increase in population cannot by itself lead economic development without increasing effective demand. Malthus rejects the Says Law, which says "supply creates its own demand". Malthus showed that excess savings in the sense of not consuming create a negative impact on economy and savings lead to a decline in effective demand. Malthus discouraged the excess savings because he feared that this tendency could reduce the effective demand. This statement is true in some cases for Bangladesh, most of the families want to save in order to protect their future and some other expenses during their retirement age. As more than 70% of people live in rural area and 24.3% population lives under the national poverty line. They don't have enough purchasing power to their subsistence. Foreign capital dependence has a positive effect on income inequality, raises fertility rates, accelerates population growth, and retards economic development (Kentor, 2000). We cannot totally avoid the necessity of domestic saving because it can secure the domestic financial market, reduce the dependency of foreign debt and government can borrow from the domestic capital market to invest them into a local project to boost the economy. In addition, savings are needed to fulfill the budget deficit and government can take comparatively low-interest rate than abroad.

**Discussion**

Theory of "Dualism" by Malthus is very true for the underdeveloped countries. Malthus suggested land reform to expand agricultural output. Most of the recently developed countries were also went through the process. Malthus explain the concept of economic development, he says economy consists of two major sectors such as an agricultural and industrial sector. Most of the underdeveloped countries have the dual economy and agricultural sector is lagging behind than the industrial sector in these countries. Law of diminishing returns operated in the agricultural sector even though technological progress in the production process, whereas the industrial sector was subject to the law of increasing returns. The most important point is that when one of these sectors lags behinds, it held back the development of another sector. In Bangladesh, this struggle agricultural sector hinders the development of the industrial sector. 70% population of Bangladesh lives in rural areas and they are solely dependent on agriculture and most of them live under the poverty line. So, they don't have enough purchasing power to reflect the exact demand. Due to the lack of purchasing power, they can't able to show the demand that reduces effective demand in the economy and retards its growth. Bangladesh has been facing serious environmental problems due to the population growth. In recent time GDP is growing rapidly with the population growth, which puts downward pressure on the country's natural resources and increases the level of pollution. The economy of Bangladesh is growing very in recent times and annual GDP growth is 7.3% in 2017. This rapid economic growth causes damage to the environment and reduces ambient quality; rapid expansion in industrial production and urbanization has caused water pollution, increased the level of industrial waste that has resulted in health problems.

Malthus's Theory of development is kind of negative because he just focuses on the elements that hinder the growth instead of the elements that promote the economic progress. However, he taking account two most influential factors production and distribution as the elements of economic progress in his Book Principle of Political Economy. In addition distribution of production is very important for the less developed country like Bangladesh.

Malthus also emphasized the non-economic factors, the most influential subjects are politics and morality. Proper Property Rights can encourage to worker works more. In addition to political stability, and proper law and enforcement are also needed for the development of an economy. Malthus didn't mention the women empowerment, which is the most crucial factor for the economic development. Another important thing should be considered for the growth of economics is the development of institutions and infrastructure. These two key elements can enhance sustainable economic growth.

**Conclusion**

Malthus contribution to economic development consists of several elements that are very true for developing countries like Bangladesh, Pakistan, and so on. Findings of this study are similar to the Malthus assumption of unchecked population growth can cause disaster for the economy. This article found a highly significant negative correlation between population growth and economic development. Excessive population growth can shrink the possibility of economic growth unless government come forward and take proper steps to reduce the food problems using alternatives methods like export skilled labor to abroad, promote local small industries, agricultural advancement programs, promote labor-intensive industries (because of availability of cheap labor), encourage foreign investors, institution settings and political stability. In addition, his emphasis on production, distribution, accumulation of capital, methods of increasing effective demand and non-economic factors are also true in some cases for Bangladesh.

Bangladesh is trying to improve its food security using different agricultural advancement programs, government subsidy in the agricultural sector and investment in research and development. Recent statistics show that Bangladesh achieved tremendous success in the agricultural sector. Agriculture is a key in reducing poverty in Bangladesh and agriculture is accounted for 93% of the reduction in poverty between the periods of 2005-2015. Agricultural growth in Bangladesh stimulates non-farm income and 10% increases in farm income generates a 6% increases in non-farm income. In addition, Bangladesh made huge progress in achieving food security ever before. This rapid increase in agricultural productivity falsified the Malthus theory of population.

Malthus only accounted the relationship between agricultural land and food production alone but he didn't consider all types of products like natural resource, export human capital, factories, industries, machinery, mines, and others specialized industries. Using the advantage of population growth Bangladesh is exporting labor to the abroad and earning huge amount of remittance and makes labor-intensive industries like textile and readymade garments. These two key things enhance food security, solve food problems and accelerate the economic development of Bangladesh. Malthus has ignored these things when he talked about economic growth.

According to recent statistics, Bangladesh economy is growing at the cost of environmental degradation. Malthus said unchecked population can cause damage and ended up misery and starvation. Malthus didn't consider the economic growth at the cost of environmental degradation and natural resource. In Bangladesh, recent GDP growth rate is 7.3% (2017), GDP per capita PPP is $3868 in 2017 (340% percent higher than the year 1990), life expectancy increased 58 years to 71 years between the period of 1990 and 2016, prenatal care for pregnant women increased 63.9 % from 25% from the period of 1994 to 2015. Some others indicators like mortality rate, reduce child dead at birth, reduction in poverty and so on indicates living standards is improving. On the other hand, the cost of economic growth is huge in terms of population growth, arable land, agricultural land, and forestland are decreasing faster rate, which is alarming for Bangladesh. If the government of Bangladesh doesn't proper steps to solve the problems, near future she will lose most of the cultivated land. Excess Carbon emission enhances climate change and reduces ambient air quality. As a result sea level rises and they have to suffer natural disasters, Bangladesh already has been going through these problems.

Malthus pointed out two important checks to reduce the growth of populations; these are preventive check and positive check. Bangladesh has taken steps to control population growth and they made huge success in controlling population, population growth reduces to 1% form 3% between the periods of 1979 to 2017. Bangladesh has taken the plan to spend $615 million USD for the family planning program over 2017-2021 to reduce population growth. The main reasons for success behind the reduction of population growth are women empowerment and education. In Bangladesh, most women don't work outside, which is the key obstacles for economic development because women are accounted for half of the workforce of the country. Women participation in the workforce is very important for the development of any country.